\documentclass[12pt,a4paper,epsf]{article}
\usepackage{graphics}
\usepackage{amssymb,amsmath}
\usepackage[dvips]{lscape,graphicx}
\usepackage{cite}
\usepackage{longtable}

\newcommand{\ct}{\cite}
\newcommand{\lb}{\label}

\newcommand{\bc}{\begin{center}}
\newcommand{\ec}{\end{center}}
\newcommand{\bd}{\begin{displaymath}}
\newcommand{\ed}{\end{displaymath}}
\newcommand{\be}{\begin{equation}}
\newcommand{\ee}{\end{equation}}
\newcommand{\ba}{\begin{array}}
\newcommand{\ea}{\end{array}}
\newcommand{\bea}{\begin{eqnarray}}
\newcommand{\eea}{\end{eqnarray}}
\newcommand{\bt}{\begin{tabular}}
\newcommand{\et}{\end{tabular}}
\newcommand{\un}{\underline}
\newcommand{\ov}{\overline}
\newcommand{\bp}{\begin{picture}}
\newcommand{\ep}{\end{picture}}
\newcommand{\bfi}{\begin{figure}}
\newcommand{\efi}{\end{figure}}

\def\fun#1#2{\lower3.6pt\vbox{\baselineskip0pt\lineskip.9pt
\ialign{$\large \bf\mathsurround=0pt#1\hfil##\hfil$\large
\bf\crcr#2\crcr\sim\crcr}}}

\begin{document}



\vspace{1cm}

\title{\Large \bf New Bound States of Heavy Quarks\\ at LHC and
Tevatron}
\author{\large\bf C.R.~Das $\large \bf{}^{1}$\large \bf
\footnote{crdas@cftp.ist.utl.pt}\,\,, C.D.~Froggatt $\large
\bf{}^{2}$\large \bf \footnote{c.froggatt@physics.gla.ac.uk}\,\,,
L.V.~Laperashvili $\large \bf{}^{3}$\large \bf
\footnote{laper@itep.ru}\,\, and
H.B.~Nielsen $\large \bf{}^{4}$\large \bf \footnote{hbech@nbi.dk}\\[5mm]
\itshape{\large $\large \bf{}^{1}$ Centre for Theoretical Particle
Physics,}\\ \itshape{ Technical University of Lisbon, Lisbon,
Portugal}\\ \itshape{ $\large
\bf{}^{2}$ Department of Physics and Astronomy,}\\
\itshape{ Glasgow University, Glasgow, Scotland}\\
\itshape{\ $\large \bf{}^{3}$ Institute of Theoretical and
Experimental Physics,}\\ \itshape{ Moscow, Russia}\\
\itshape{$\large \bf{}^{4}$ The Niels Bohr Institute, Copenhagen,
Denmark }}

\date{}

\maketitle

\thispagestyle{empty}

\vspace{1cm}

\begin{abstract}
The present paper is based on the assumption that heavy quarks
bound states exist in the Standard Model (SM). Considering  New
Bound States (NBS) of top-anti-top quarks (named T-balls) we have
shown that: 1) there exists the scalar $1S$--bound state of
$6t+6\bar t$; 2) the forces which bind the top-quarks are very
strong and almost completely compensate the mass of the twelve
top-anti-top-quarks in the scalar NBS; 3) such strong forces are
produced by the Higgs-top-quarks interaction with a large value of
the top-quark Yukawa coupling constant $g_t\simeq 1$. Theory also
predicts the existence of the NBS $6t + 5\bar t$, which is a color
triplet and a fermion similar to the $t'$-quark of the fourth
generation. We have also considered the ``b-quark-replaced'' NBS,
estimated the masses of the lightest fermionic NBS:
$M_{NBS}\gtrsim 300$ GeV, and discussed the larger masses of
T-balls. We have developed a theory of the scalar T-ball's
condensate and predicted the existence of three SM phases.
Searching for heavy quark bound states at the Tevatron and LHC is
discussed. We have constructed the possible form-factors of
T-balls, and estimated the charge multiplicity coming from the
T-ball's decays.
\end{abstract}

\vspace{1cm}

\section{Introduction}

Although the Standard Model (SM) was confirmed by all experiments
of the world accelerators, the mechanism of the Electroweak (EW)
symmetry breaking (EWSB) has not yet been tested. According to the
SM, the Higgs boson is responsible for generating the masses of
fermions due to the  Higgs mechanism. However, the mass of the
Higgs boson is not predicted by theory.

Direct searches in the previous experiments (mainly at LEP2\,
\ct{1}) set a lowest limit for the Higgs boson mass $M_H$:
\be M_H \gtrsim 114.4 \,\,\, GeV\,\,\, at \,\,\, 95\% \,\,\, CL.
\lb{1a} \ee
The recent Tevatron result \ct{2} is:
\be 114 \lesssim M_H \lesssim 158\,\,\, GeV \lb{2a} \ee
at 97\% C.L. if direct limit of 114 GeV from LEP is excluded in
thw fit. We hope that LHC will provide a solution of main puzzles
of EWSB.

The Higgs boson couples more strongly to the heavy top quarks than
to the light ones. As a result, the Higgs exchanges between top
quarks produce new type of bound states
\ct{3,3a,4,5,6,7,8,9,10,11,12,13,14,15}.

The present paper is devoted to the properties of the new bound
states (NBS): estimates of their masses and observation at modern
colliders (Tevatron, LHC, etc.). The predictions of
Refs.~\ct{3,3a,4,5,6,7,8,9,10,11} are:

$\bullet$ There exists a scalar $1S$--bound state of $6t + 6\bar
t$. The forces which bind these top-quarks are so strong that
almost completely compensate the mass of the 12 top-quarks forming
this bound state.

$\bullet$ There exists a new bound state $6t + 5\bar t,$ which is
a fermion similar to the quark of the fourth generation having
quantum numbers of top quark.

$\bullet$ Theory also predicts the existence of new bound states
with b-quark replaced the t-quark: for example, NBS $ n_b b + (6t
+ 6\bar t - n_b t)$, etc., where $ n_b=1,...6.$

A new (earlier unknown) bound state $6t+6\bar t,$ which is a color
singlet (that is, `white' state), was first suggested by Froggatt
and Nielsen in Ref.~\ct{4}. Now all these NBS are named T-balls,
or T-fireballs.

\section{Higgs and gluon interactions of quarks}

If the Higgs particle exists, then between quarks $qq$, quarks and
anti-quarks $q\bar q$, and also between anti-quarks $\bar q\bar q$
there exist virtual exchanges by Higgs bosons (see Fig.~1),
leading only to the attractive forces.

 \bfi \centering
\includegraphics[height=80mm,keepaspectratio=true,angle=0]{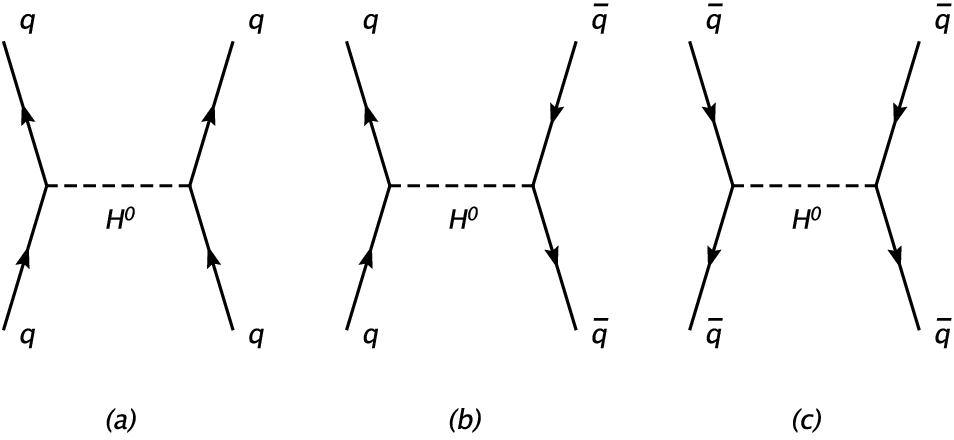}\caption{}\efi

It is well-known that the bound state $t\bar t$ -- so called
toponium -- is obliged to the gluon virtual exchanges of Fig.~2.
Among a considerable quantity of articles devoted to the toponium,
we distinguish the following backward papers
\ct{16,17,18,19,20,21}.

\bfi \centering
\includegraphics[height=80mm,keepaspectratio=true,angle=0]{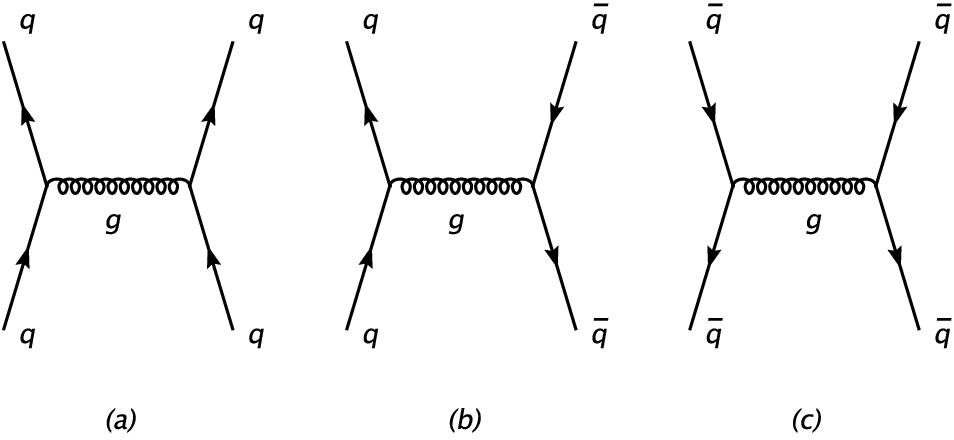}\caption{}\efi

In the case of the toponium the contributions of the Higgs scalar
particles are essential, but less than gluon interactions.
Toponium is very unstable due to the decay of the top quark
itself. However, putting more and more top and anti-top quarks
together in the lowest energy bound states, we notice that the
attractive Higgs forces continue to increase. Simultaneously gluon
(attractive and repulsive) forces first begin to compensate
themselves, but then begin to decrease relatively to the Higgs
effect with growth of the number of top-anti-top constituents in
the NBS.

The maximum of the binding energy value corresponds to the
$1S$-wave state of the NBS $6t + 6\bar t$. The explanation is
simple: top-quark has two spin states and three states of colors:
$2\times 3=6$ degrees of freedom. This means that, according to
the Pauli principle, only 6 pairs of $t\bar t$ can simultaneously
exist in the `white' $1S$-wave state. If we try to add more $t\bar
t$-pairs , then some of them will turn out to the $2S$-wave state,
and the NBS binding energy will decrease at least 4 times. For
P-,D-, etc. wave states the NBS binding energy decreases more and
more.

\section{T-ball mass estimate}

The kinetic energy term of the Higgs field and the top-quark
Yukawa interaction are given by the following Lagrangian density:
\be L = \frac 12 D_{\mu}\Phi_H D^{\mu}\Phi_H + \frac{g_t}{\sqrt
2}\ov{\psi_{tL}}\psi_{tR}\Phi_H  + h.c.,   \lb{1} \ee where
$\Phi_H$ and $\psi_t$ are the Higgs and top-quark fields,
respectively, and $g_t$ is the Yukawa coupling constant of their
interaction.

The VEV of the Higgs field in the EW-vacuum is:
\be v=<|\Phi_H|>=246\,\, {\rm{GeV}}.\lb{2} \ee
According to the Salam-Weinberg theory the top-quark mass $M_t$
and the Higgs mass $M_H$ are given by the following relations:
\be  M_t = \frac {g_t}{\sqrt 2}v \quad {\rm{and}} \quad M_H^2
=\lambda v^2, \lb{3} \ee
where $\lambda$ is the Higgs self interaction coupling constant.

According to Ref.~\ct{22},
\be M_t\approx 172.6\,\,{\rm{ GeV}},\lb{4} \ee
and
\be g_t \approx 0.93. \lb{5} \ee
Let us imagine now that the NBS is a bubble in the EW-vacuum and
contains $N_{const.}$ top-like constituents. It is known that
insight the bubble (bag) the Higgs field can modify its VEV.
Implications related with this phenomenon have been discussed in
Refs.~\ct{5,21,23,24,25,26,27}. Then insight T-balls the VEV of
the Higgs field is smaller than $v$:
\be v_0 = <|\Phi_h|>,\quad {\rm{where}}\quad \frac{v_0}{v} <
1,\lb{6} \ee
and the effective masses insight the bubble (bag) are smaller than
the corresponding experimental masses:
\be m_{t,h} = \frac{v_0}{v}M_{t,H}.\lb{7} \ee
In this case the attraction between two top (or anti-top) quarks
is presented by the Yukawa type of potential:
\be  V(r) = - \frac{g_t^2/2}{4\pi r}\exp(-m_hr). \lb{8} \ee
Assuming that the radius $R_0$ of the bubble is small:
\be m_hR_0 << 1,\lb{9} \ee
we obtain the Coulomb-like potential:
\be V(r) \simeq - \frac{g_t^2/2}{4\pi r}. \lb{10} \ee
The attraction between any pairs $tt,\,\,t\bar t,\,\,\bar t\bar t$
is described by the same potential (\ref{10}).

By analogy with Bohr Hydrogen-atom-like model, the binding energy
of a single top-quark relatively to the nucleus containing
$Z=N_{const.} - 1$ top-quarks have been estimated in
Refs.~\ct{4,5,6}. The total potential energy for the NBS with
$N_{const.}= 12$ is:
\be  V_{tot}(r) = - 11\frac{g_t^2/2}{4\pi r}. \lb{11} \ee
Here we would like to comment that the value of the mass $m_h$,
which belongs to the Higgs field insight the NBS $6t + 6\bar t$,
can just coincide with estimates given by Refs.~\ct{12,13,14,15}.
The results: $\rm max (m_h)=29$ Gev and $\rm max (m_h)=49$ Gev
correspond to Ref.~\ct{13} and Ref.~\ct{15}, respectively.

Considering a set of Feynman diagrams (the Bethe-Salpeter
equation) and including the contributions of all (s-,t- and u-)
channels for the Higgs and gluon exchange forces (see Ref.~[6]),
we obtain the following Taylor expansion:
\be
    M_T^2 = (N_{const.}M_t)^2\times \left\{1 -
     2(N_{\mbox{const.}}-1){\left(\frac{
N_{const.}}{12}\right)}^2\left(\frac{g_t^2 + \frac 16
g_s^2}{\pi}\right)^2 +....\right\}. \lb{12} \ee
Here the QCD coupling constant $g_s$ is given by its fine
structure constant value at the EW-scale \ct{22}:
\be
       \alpha_s(M_Z) = g_s^2(M_Z)/4\pi \approx 0.118. \lb{14} \ee
Now the value of the total binding energy for arbitrary
$N_{const.}$ is equal to:
 \be
    E_T = N_{const.}(N_{const.} - 1){\left(\frac{ N_{const.}}{12}\right)}^2
    {\left(\frac{g_t^2 + \frac 16 g_s^2}{\pi}\right)}^2
    m_t. \lb{15} \ee
The mass of T-ball containing $N_{\mbox{const.}}$ top or anti-top
quarks is:
 \be  M_T = N_{\mbox{const.}}m_t - E_T.     \lb{16} \ee

Approximately this dependence is described by the following
expression:
 \be   M_T = N_{\mbox{const.}}m_t\left\{1 -
(N_{\mbox{const.}}-1){\left(\frac{
N_{const.}}{12}\right)}^2\left(\frac{g_t^2 + \frac 16
g_s^2}{\pi}\right)^2\right\}.   \lb{17} \ee

Below we shall use the following notations: $T_s$-ball is a scalar
NBS $6t + 6\bar t$, having the spin $S=0$, and $T_f$-ball presents
the NBS $6t + 5\bar t$, which is a fermion: $\ov {T_f} = 5t +
6\bar t$.

Let us consider now the condition:
\be  \frac{11}{\pi^2}\cdot (g_t^2 + \frac 16g_s^2)^2 = 1. \lb{18}
\ee
In this case  the binding energy $E_T$ compensates the NBS mass
$12 m_t$ so strongly that the mass of the scalar $T_s$-ball
becomes zero:
 \be  M_{T_s} = 11m_t\left\{1- \frac{11}{\pi^2}\cdot (g_t^2 + \frac 16g_s^2)^2\right\}=
 0. \lb{19} \ee
It is necessary to emphasize that the experimental values given by
(\ref{5}) and (\ref{14}) \ct{22}:
\be g_t^2\simeq 0.86 \quad {\rm {and}} \quad g_s^2\simeq 1.48
\lb{20} \ee
are just very close to this limit.

Fig.~3 shows the dependence of T-ball masses on the number of NBS
constituents $N_{\mbox{const.}}$. In the case when $M_{T_s} = 0$,
we have:
\be M_T = N_{\mbox{const.}}m_t\left\{1 -
\frac{(N_{\mbox{const.}}-1)}{11}\frac{N_{const.}^2}{12^2}\right\}.
\lb{21} \ee

\bfi \centering
\includegraphics[height=100mm,keepaspectratio=true,angle=0]{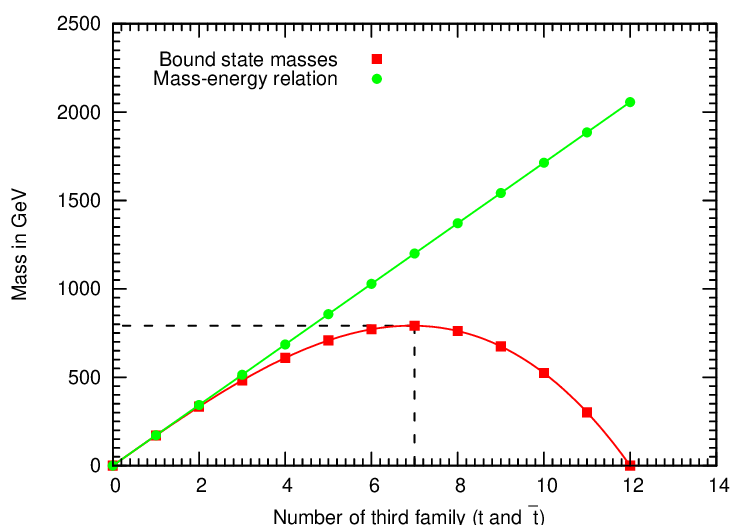}
\caption{T-ball mass depending on the number $N_{\mbox{const.}}$
of the NBS constituents. } \lb{4f}\efi

We easily see that the light scalar Higgs bosons with mass $m_h <
M_H$ can bind the 12 top-like quarks so strongly that the mass
$M_{T_s}$ becomes almost zero, and even tachyonic: $M_{T_s}^2 <
0$. In the last case we obtain the Bose-Einstein condensate of
T-balls -- a new vacuum at the EW-scale \ct{11}. Previously the
condensation of $t\bar t$, arising from four-fermion interaction
models (\ct{28,29,30}, etc.), was reviewed in Ref.~\ct{31}. We
have suggested a new type of condensation of top-quarks via
T-balls, what is very important for the solution of the hierarchy
problem in the SM \ct{9,10}.

\subsection{$\large \bf T_f$-ball mass estimate}

As we have discussed above, the Higgs interaction of the eleven
top-anti-top quarks ($N_{\mbox{const.}}=11$) creates a $T_f$-ball
-- a new fermionic bound state $6t + 5\bar t$, which is similar to
the $t'$-quark of the fourth generation. The estimate of the mass
of $T_f$-ball $6t + 5\bar t$ by Eq.~(\ref{21}) gives :
\be  M_{T_f}\approx 11m_t\cdot 0.236\,\,\gtrsim\,\, 300\,\, GeV.
\lb{22} \ee
The detailed analysis of calculation of the NBS-masses was
considered in Ref.\ct{6}. We hope that the forthcoming numerical
calculations of the T-ball masses by Monte-Carlo simulations on
lattice will give us more exact answers.

\section{New ``b-replaced'' bound states}

Constructing T-balls from $t$ and $\bar t$-quarks, we also can
take into account considerable contributions of left b-quarks
insight NBS \ct{3,6,11}.

If we had no $b\bar b$-pairs in T-balls, then there would be an
essential superposition of different states of the weak isospin.
The presence of b-quarks in the NBS leads to the dominance of the
isospin singlets of EW-interactions only. Now such a
``b-replaced'' scalar NBS would be stable. We predict the
following scalar ``b-replaced'' NBS:
\be T_s(b-replaced) = b + 5t + 6\bar t, \lb{B1} \ee
\be T_s(\bar b-replaced) = 6t + \bar b + 5\bar t. \lb{B2} \ee
In general case we can construct the following scalar
``b-replaced'' T-balls:
\be T_s(nb-replaced) = n_b b + (6t + 6\bar t - n_b t), \lb{B3} \ee
and
\be T_s(n\bar b-replaced, ) = n_{\bar b}\bar b + (6t + 6\bar t -
n_{\bar b}\bar t). \lb{B4} \ee
Of course, we also can construct the fermionic ``b-replaced''
NBS:
\be T_f(b-replaced)= b + 5t + 5\bar t, \lb{B5} \ee
and
\be \ov{T_f}(\bar b-replaced)= 5t + 5\bar t + \bar b. \lb{B6} \ee
In general case we obtain:
 \be T_f(nb-replaced) =
n_b b + (6t + 5\bar t - n_b t), \lb{B7} \ee
and
\be \ov{T_f}(n\bar b-replaced) = n_{\bar b}\bar b + (5t + 6\bar t
- n_{\bar b}\bar t). \lb{B8} \ee
We have $n_b, n_{\bar b}=1,...6$ in Eqs.~(\ref{B3})-(\ref{B8}).

 There is a simple way to estimate the mass of the
``b-replaced'' T-ball with one t-quark replaced by a b-quark. It
is well-known that b-quark does not interact significantly with
NBS. Thus, we can add a b-quark (or anti-b-quark) to the NBS
having eleven constituents without essential changing its energy,
or mass. Then the b-replaced scalar NBS $T_s(b-replaced)$, or
$T_s(\bar b-replaced)$, given by Eqs.~(\ref{B1}) and (\ref{B2})
respectively, will have a mass $\,\,\backsimeq 300$ GeV.

As to the NBS $T_f(b-replaced) = 5t + b + 5\bar t$ and
$T_f(b-replaced, b\bar b) = 5t + b + n_b b\bar b + 5\bar t$, they
will have a mass very close to the NBS with ten constituents, e.g.
$M_{T_f} \backsimeq 500$ GeV (see Fig.~3).

We also can consider more heavy T-balls with $M_T > 500$ GeV, but
they will have very small cross-sections of their production.

The more accurate estimate given in Ref.~\ct{6} predicts the
existence of "11" and "10" constituent bound states with masses
approximately 760 and 960 GeV, respectively.

\section{New phases of the SM}

The existence of the new phases of the SM, different from the
well-known Salam-Weinberg Higgs phase, leads to the confrontation
with a question: Does a phase of the condensed $T_s$-balls exist?

The answer on this question is related with the SM parameters.

We can consider two phases I and II:

Phase-I does not have the Bose-Einstein condensate of $T_s$-balls.
In this phase the VEV of the $T_s$-ball's scalar field $\Phi_T$ is
equal to zero: $<\Phi_T>=0$.

Phase-II contains such a condensate and $<\Phi_T>\neq 0$.

The main requirement of the appearance of the new phase of the
condensed $T_s$-balls is a condition:
$$m^2_{NBS} = M_{T_s}^2 = 0.$$

\subsection{Three EW phases of the SM}

\bfi \centering
\includegraphics[height=120mm,keepaspectratio=true,angle=0]{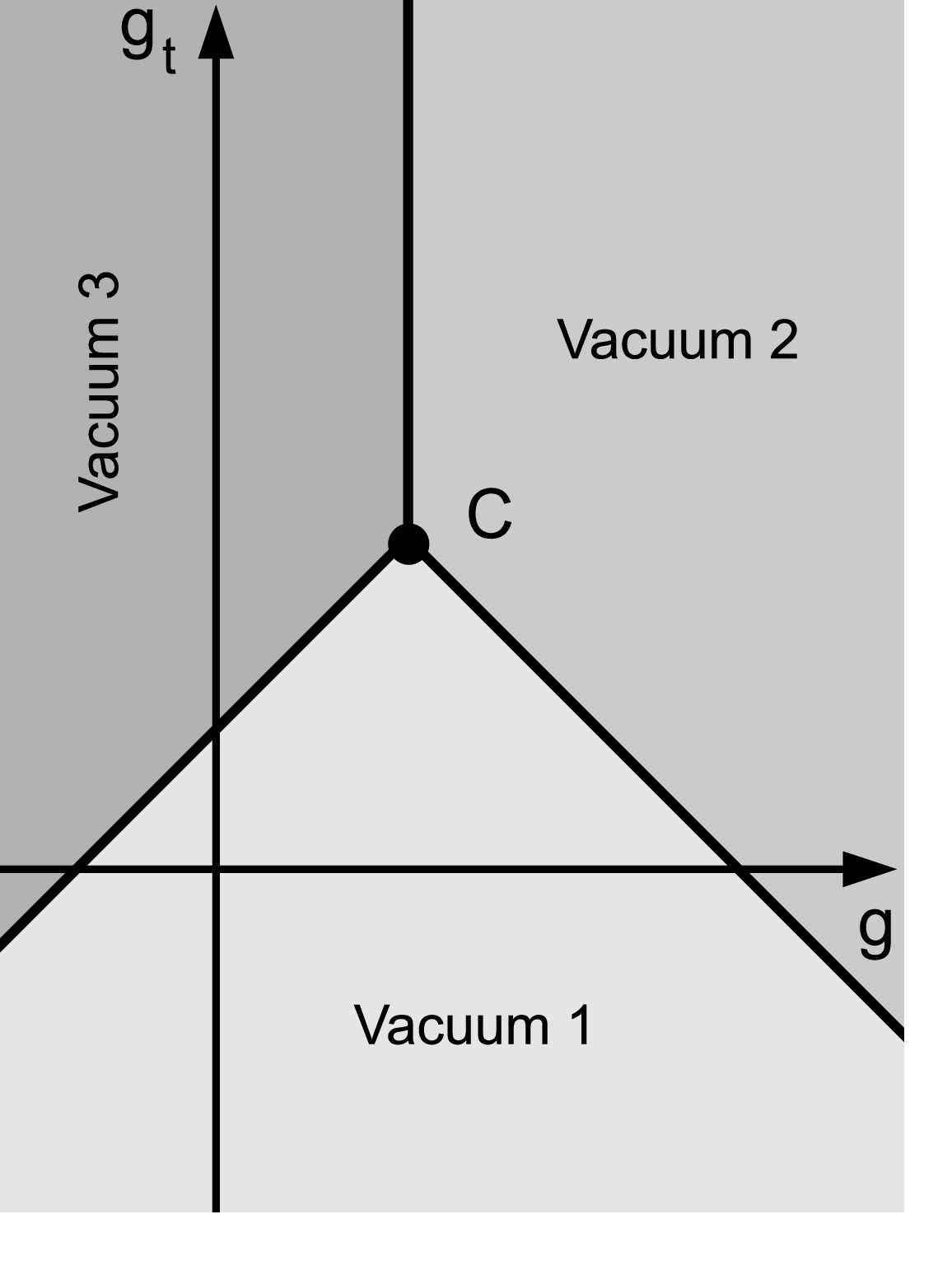}
\caption{A symbolic phase diagram for the SM at the EW-scale.}\efi

Finally, taking into account seriously our results in the
estimates of $g_t$ and $M_T$, we can consider three phases --
three vacua of the SM at the EW-scale:\\

I) $<\Phi_H> \neq 0 , <\Phi_T> = 0$ --- ''Vacuum 1'', the phase
in which we live;\\

II) $<\Phi_H> \neq 0 , <\Phi_T> \neq 0$ --- ''Vacuum 2'';\\

III) $<\Phi_H> = 0$,
$<\Phi_T> \neq 0$ --- ''Vacuum 3'',\\
which are presented symbolically by the phase diagram of Fig.~4.

Fig.~4 shows the critical point C (triple point), in which the SM
three phases meet together: this triple point is similar to the
critical point considered in thermodynamics where the density of
the vapor, water and ice are equal (see Fig.~5).

\bfi \centering
\includegraphics[height=60mm,keepaspectratio=true,angle=0]{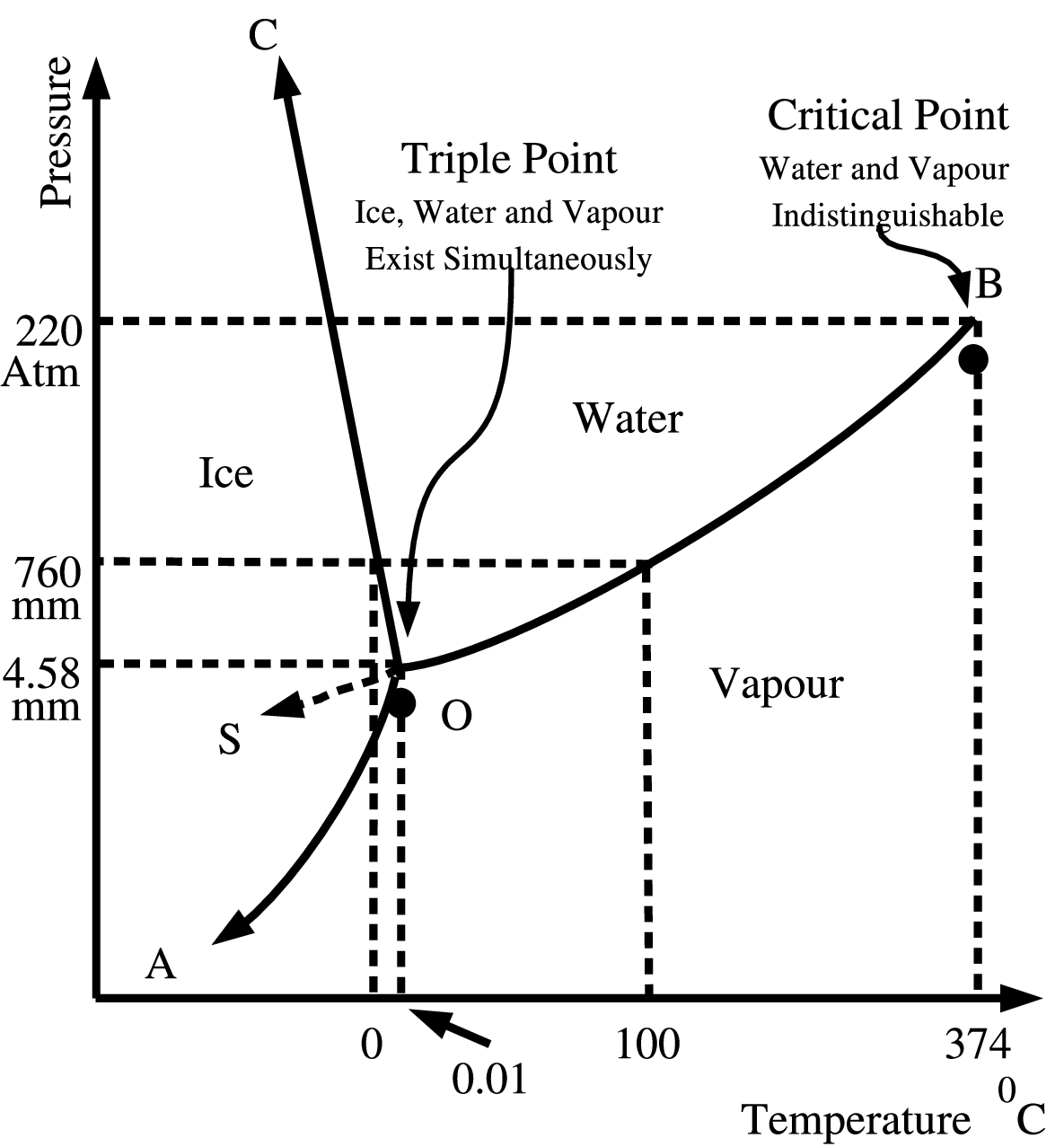}\caption{}\efi

The existence of the new phases near the EW-scale can solve the
problem of hierarchy. Here we recall the Multiple Point Principle
(MPP) suggested in Refs.~\ct{10z,11z,12z,13z,14z,15z,16z,17z}.

\subsection{The fundamental (Planck) scale of the SM}

{\it A priori} it is quite possible for a quantum field theory to
have several minima of its effective potential as a function of
its scalar fields $\Phi$ (exactly speaking of its norm $|\Phi|$).
These minima can be degenerate. Moreover, it is assumed that all
vacua existing in Nature (there can be a number of several vacua)
are degenerate and have the same zero, or almost zero, vacuum
energy densities which coincide with the cosmological constant
$\Lambda$ determined by Einstein. This is confirmed by the
phenomenological cosmology.

According to the MPP, the SM has the two minima of its effective
potential considered as a function of the variable $|\Phi|$, where
$\Phi=\Phi_H$. These minima are degenerate and have $\Lambda = 0$:
\be
              V_{\mbox{eff}}|_{min1}= V_{\mbox{eff}}|_{min2}=0,
\lb{11} \ee \be  \bf
             V'_{\mbox{eff}}|_{min1}= V'_{\mbox{eff}}|_{min2}=0,   \lb{12z}
\ee what is shown in Fig.~6.

\bfi \centering
\includegraphics[height=80mm,keepaspectratio=true,angle=0]{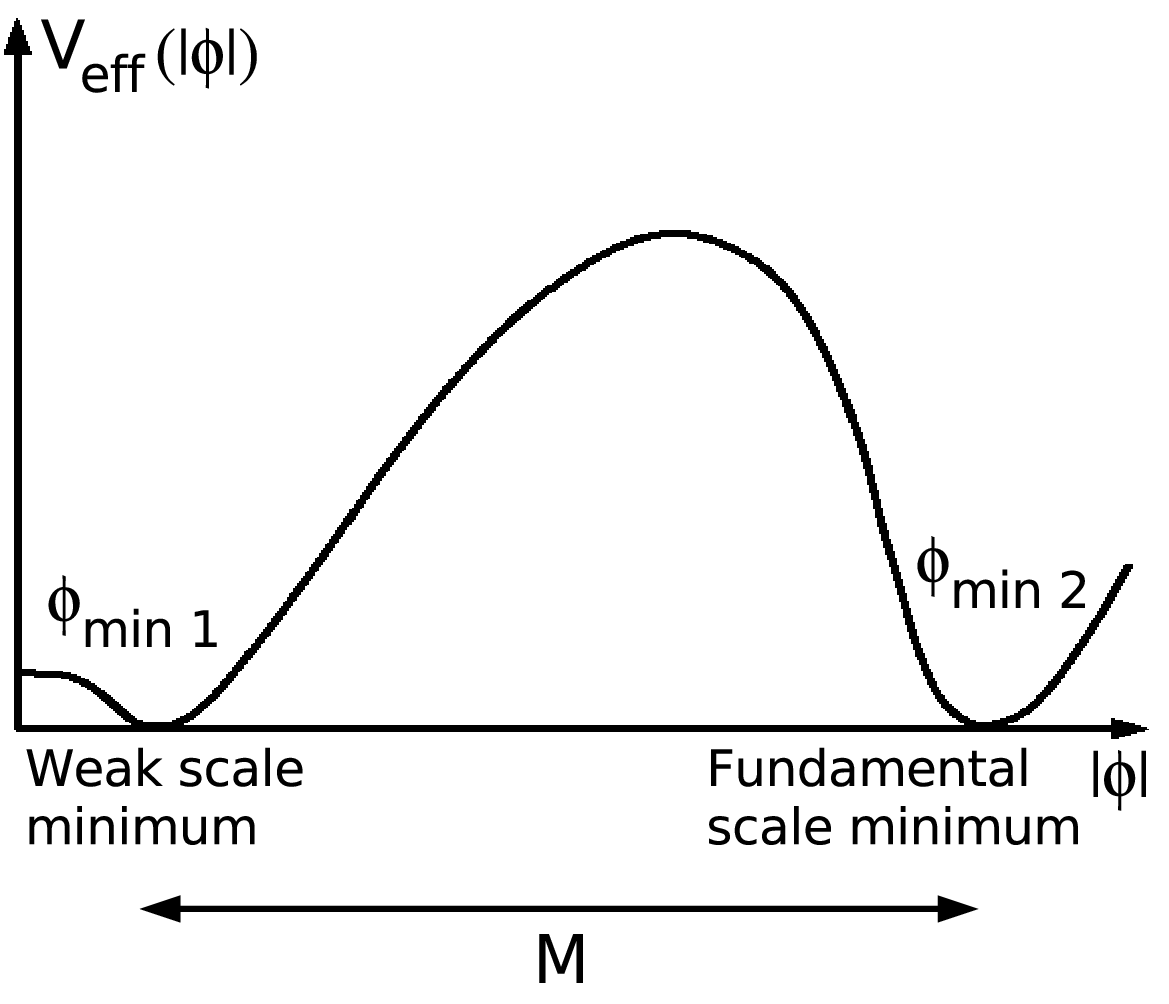}
\caption{The first (our) vacuum at $|\Phi|\approx 246$ GeV and the
second vacuum at the fundamental scale $|\Phi|\sim
M_{\mbox{Pl}}$.}\efi

It is assumed that the second minimum exists near the Planck
scale:
$$ |\Phi_{min2}|\sim M_{\mbox{Pl}}.$$
This is a fundamental scale.

The calculation of the NBS masses is based only on the SM
parameters. The MPP determines the coupling constants in the SM
and therefore
--- the structure of the NBSs $T_{s,f}$. Since at the border of
the two phases I and II the top-quark Yukawa coupling constant
leads to zero mass of the NBS $T_s$, we can assume that the MPP
manifests the phase transitions in the SM in such a way that we
have the finetuning in the SM, which solves the hierarchy problem.
The MPP calculations of gauge coupling constants were obtained in
Refs.~\ct{18z,19z}.

\section{Can we observe T-balls at LHC or Tevatron?}

If the mean square radius of the T-ball is small in comparison
with its Compton wave length:
\be  r_0\approx (\sqrt 2 M_t)^{-1} << \frac{1}{m_{NBS}},\lb{r0}
\ee
then the NBS can be considered as an almost fundamental particle.

The fermionic NBS $T_f$ is similar to the $t'$-quark of the fourth
generation belonging to the fundamental representation {\un 3}
(color triplet).

Then our NBS are strongly bound and can be observed at colliders
 (Tevatron, LHC, etc.) in the following
processes:

1) First of all, in the possible H-decay process:
 \be H\to 2T_s, \lb{23z} \ee
if $M_{T_s} < \frac 12 M_H.$ Using limits given by Tevatron
experiments \ct{2}: $114 \lesssim M_H \lesssim 158\,\,\, GeV,$ we
obtain the requirement for the Higgs decay mechanism:
\be M_{T_s} \lesssim 80 \,\, GeV.  \lb{24z} \ee
Here we have argued that T-balls can explain why it is difficult
to observe the Higgs boson H at colliders: T-balls can strongly
enlarge the decay width of the Higgs particle.

2) If $M_{T_s} > \frac 12 M_H$, then the first decay (\ref{23z})
is absent in Nature, and the $fT_s$-balls fly away, forming jets
which produce hadrons with a high multiplicity:
\be T_s \to JETS. \lb{25z} \ee
3) Second, we can observe at Tevatron all processes given by
Fig.~7 with the replacement $t\bar t \to t'\bar
t',\,\,T_f\ov{T_f}$. In the most optimistic cases the NBS $6t +
5\bar t$ (fermionic fireball) plays a role of the fundamental
quark of the fourth generation, say, with the mass $M_{T_f}\gtrsim
300$ GeV, given by our preliminary estimate. We expect that the
Tevatron-LHC experiments should find either a fourth family
t'-quark, or the fermionic NBS $T_f $, or both of them.

\bfi \centering
\includegraphics[height=120mm,keepaspectratio=true,angle=0]{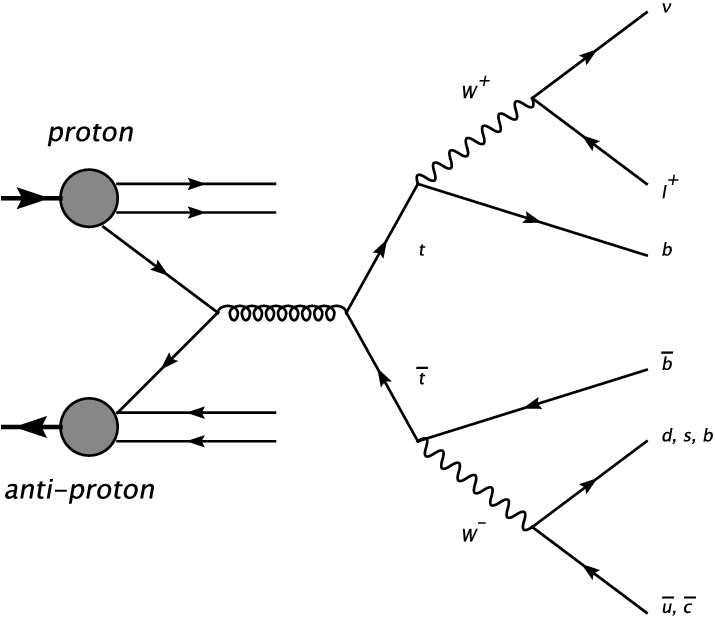}
\caption{A typical process observed at the Tevatron in $p\bar p$
collisions.}\efi

The scalar NBS $T_s$ cannot be produced simply in a pair by a
gluon vertex, because it is a color singlet $\bf \un{1}$. But a
pair $T_f\ov{T_f}$ can be produced by a gluon, because $T_f$ is a
color triplet $\bf \un{3}$.

At LHC the pairs of $T_s$-balls, or $T_f$-balls might be produced
in $pp$ collisions via the two gluon diagram with strong vertices
shown in Fig.~8 \ct{3,32}.

\bfi \centering
\includegraphics[height=70mm,keepaspectratio=true,angle=0]{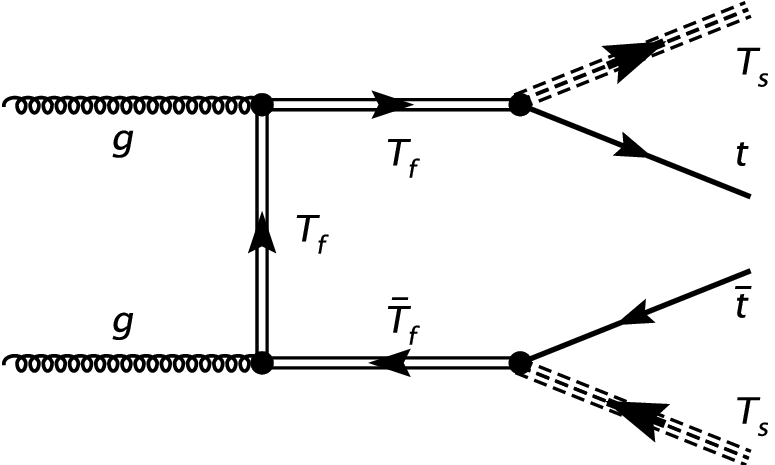}
\caption{Two gluon production of $T_s$
-balls} \efi

\section{CDF II Detector experiment at the Tevatron}

Recent experiments with CDF II Detector of the Tevatron \ct{1}
searching for heavy top-like quarks in $p\bar p$-collisions with
$\sqrt s \backsimeq 1.96$ TeV do not exclude the existence of
T-balls with masses $\gtrsim 300\,\,{\mbox{GeV}}$ up to 500 GeV.

Here we can assume that the very strange events observed at the
Tevatron as a fourth family $t'$, which decays into a $W$-boson
and a presumed quark-jet, might find another explanation in our
model: maybe it is a decay of T-balls into a $W$-boson and a gluon
jet.

Tevatron experiments exclude a fourth-generation t' quark with a
mass below 300 GeV (see Refs.~\ct{1}).  Assuming that fourth
generation $t'$-quarks does not exist in Nature, but only the
pairs of fermionic NBS $T_f$ are produced at the Tevatron, we can
give an explanation of the observed cross-sections shown in
Fig.~9.

\bfi \centering
\includegraphics[height=120mm,keepaspectratio=true,angle=0]{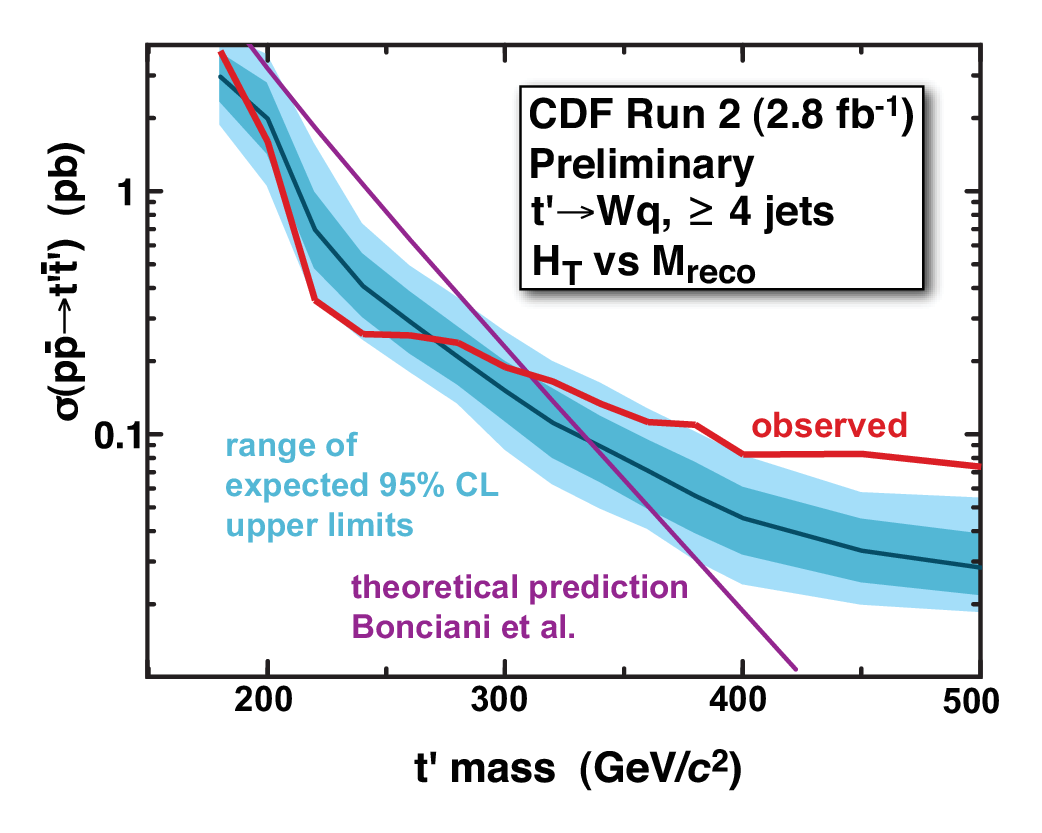}
\caption{Tevatron CDF-experiment given by Refs.~\ct{1}: upper
limit, at 95\% CL, a fourth-generation t' quark with a mass below
300 GeV is excluded. Blue line presents a theoretical curve for
the fourth-generation quarks cross-section.}\efi

The curve for the cross-section
\be \sigma(p\bar p \to t'\bar t')\simeq 0.1\,\, pb \lb{26z} \ee
can correspond to the production of pairs of fermionic $T_f$-balls
with mass $M_{T_f} \gtrsim 300$ GeV.

\section{Estimate of the NBS form-factors in the Tevatron CDF-experiment}

Assuming that only the fermionic $T_f$-balls with mass $M_T
> 300$ GeV are produced at the Tevatron in the CDF-experiment \ct{1},
we can imagine the existence of form-factors of the NBS $T_f$,
which determine the cross-section of the production of the
fermionic T-balls (see Fig.~9):
\be \sigma(p\bar p\to T_f\ov{T_f})=F^2(M_T)\sigma_{theor}(M_T).
\lb{1ff} \ee
Here $\sigma(p\bar p\to T_f\ov{T_f})$ is given by the observed red
line curve of Fig.~9 and $\sigma_{theor}(M_T)$ is given by the
theoretical (blue) curve obtained by Bonciani et al. \ct{33,34}
for the point-like particle $t'$. Our numerical calculations of
the form-factor shown in Fig.~10 gives the results in the region
of $M_T$ from 311 GeV (where $F(M_T)=1$) up to 500 GeV. We
conclude that for $M_T=500$ GeV the form-factor is large enough:
\be
       F(M_T)\approx 7.6. \lb{2ff}
\ee

\bfi \centering
\includegraphics[height=100mm,keepaspectratio=true,angle=0]{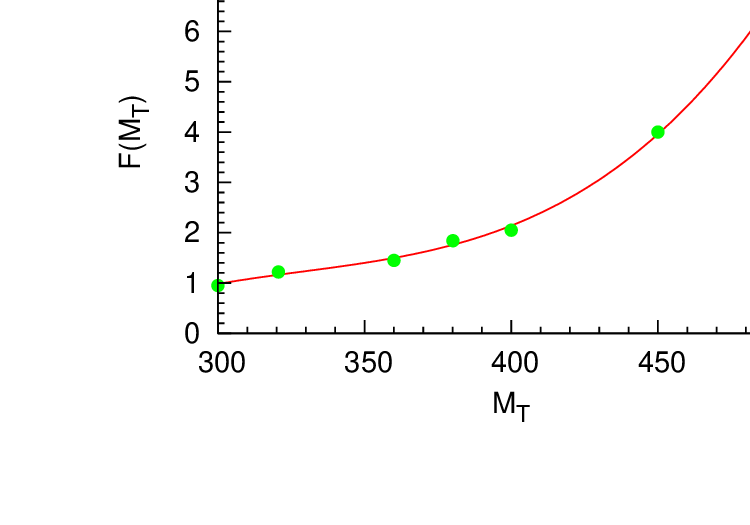}
\caption{The form-factor $F(M_T)$ of the fermionic new bound state
$T_f$ obtained from Tevatron CDF-experiment \ct{1} in absence of
the four generation.}\efi

\section{Charge multiplicity in decays of T-balls}

Actually Li and Nielsen suggested in Ref.~\ct{32} that the NBSs
would decay to a rather low number of jets, but at first one might
very reasonably think that since we have to do with bound states
of very many constituents and actually $6t\bar t$ pairs, it sounds
that the possibility of them decaying into as many jets as there
are pairs to annihilate, say - or even the number of constituents
- has some intuitive appeal and should not just be thrown away as
a possibility by the Li-Nielsen rather non-safe argument. We shall
therefore here develop what we would expect in the case of the
separate $t\bar t$ pairs decaying essentially separately, although
we do not really believe that any longer: if the mass of the NBS,
containing 6 pairs of $t\bar t$, is $M_S$, then the energy per one
annihilation of $t\bar t$ approximately is equal to the following
value:
\be   E_{an}= E_{(for\,\, one\,\, annihilation)}\approx \frac 16
M_S, \lb{15ch} \ee
e.g.
$$  E_{(for\,\, one\,\, annihilation)}\approx 10\,\,
GeV,$$ if $$ M_S\approx 60\,\,{\mbox{GeV}}.$$ In this case, during
the annihilation produced by $ e^+e^-$-collisions, the special
charge multiplicity is
$$  <N_{ch}(e^+e^-)>\approx 10,$$ while the annihilation produced
by $pp$-collisions, the special charge multiplicity is
$$<N_{ch}(pp)>\approx 6.$$

Such calculations of $<N_{ch}>$ vs $E_{an}$ are based on the
investigation of Ref.~\ct{35}. Here for $M_S\approx 60\,\, GeV$ we
obtain the following values for the charge multiplicity:
\be
    N_{ch}(e^+e^-)\approx 6\cdot 10\approx 60, \lb{16ch} \ee
\be
         N_{ch}(pp)\approx 6\cdot 6\approx 36. \lb{17ch} \ee
The value of the charge multiplicity weakly depends on the NBS
mass. For instance, if $M_S\approx 80\,\, GeV$, then:
$$ <N_{ch}(pp)>\approx 6.5,$$ and
\be
         N_{ch}(pp)\approx 6\cdot 6.5\approx 39. \lb{18ch} \ee
But if $M_S\approx 100\,\, GeV$, then:
$$ <N_{ch}(pp)>\approx 7,$$ and
\be  N_{ch}(pp)\approx 6\cdot 7\approx 42. \lb{19ch} \ee
However, such a maximally possible charge multiplicity will not be
realized in practice, because between the produced in the final
state pairs $t\bar t$, or $ b\bar b$, can exist extra exchanges by
gluons and the Higgs bosons giving new annihilations. And we shall
obtain less jets.

Indeed, it would be very strange if the decay width of the T-balls
was small. Then we would have narrow peaks in JETS. It would be
exactly a good way to see that our model were right if you could
find some narrow peak in the distribution of the total mass of
some JETS.

For $pp$-collisions the estimates \ct{32} give :
\be \frac{dN_{ch}}{d\eta}|_{max}\approx 6. \lb{20ch} \ee
Such a value is expected for this derivative at LHC (see Fig.~11).
The maximum of this curve corresponds to the LHC energy $W=14
\,\,{\mbox{TeV}}$ in $ pp$-collisions.

\bfi \centering
\includegraphics[height=100mm,keepaspectratio=true,angle=0]{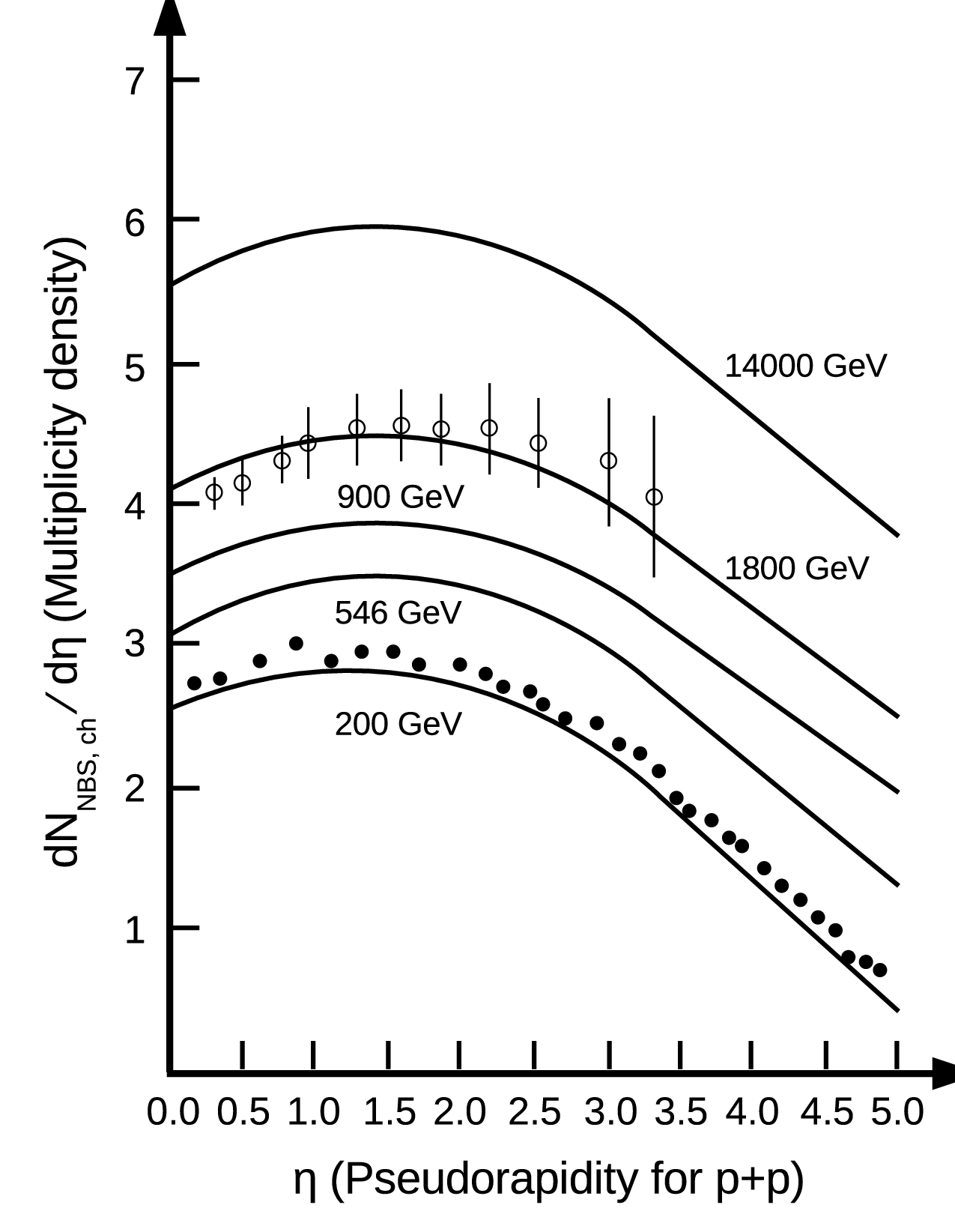}\caption{} \lb{}\efi

\clearpage \newpage

The dependence $ N_{{\mbox{ch}}}$ vs $W$ is presented in Fig.~12.
Here
\be   {N_{ch}(pp)|}_{W=14\,\, TeV}\approx 65. \lb{21ch} \ee
\bfi \centering
\includegraphics[height=100mm,keepaspectratio=true,angle=0]
{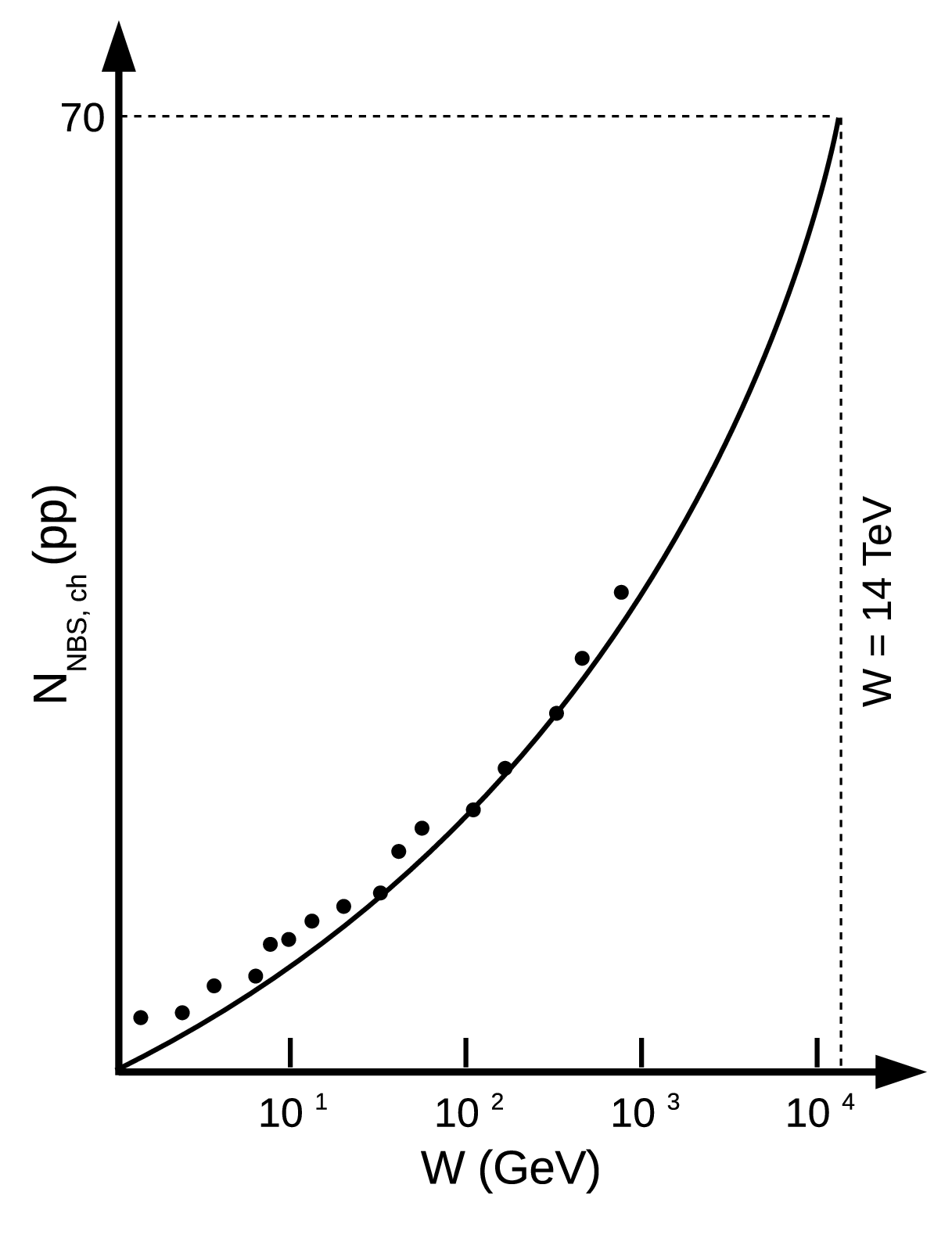}\caption{} \lb{}\efi

These calculations (figures) show that T-balls can give an
essential contributions to charge multiplicity in $pp$-collisions,
provided that their decays really go as if each $t\bar t$ pair
decayed separately and not as the recent estimate by Li and
Nielsen \ct{32}.

\section{Conclusions}

At present, a lot of physicists, theorists and experimentalists,
are looking forward to the New Physics. However, it is quite
possible that LHC will discover only the Salam-Weinberg Higgs
boson and nothing more. Nevertheless, the T-balls considered in
the present paper could exist in the framework of the SM.

1. The present investigation devoted to the main problems of the
Standard Model is based on the following three assumptions: 1)
there exists $1S$--bound state of $6t+6\bar t$, e.g. bound state
of 6 quarks of the third generation with their 6 anti-quarks; 2)
the forces which bind these top-quarks are so strong that they
almost completely compensate the mass of the 12 top-quarks forming
this bound state; 3) such strong forces are produced by the Higgs
interactions: the interactions of top-quarks via the virtual
exchange of the scalar Higgs bosons coupling with a large value of
the top-quark Yukawa coupling constant.

A new bound state $6t+6\bar t$, which is a color singlet, was
first suggested by Froggatt and Nielsen and now is named 'T-ball'.

2. Present theory also predicts the existence of a new bound state
$6t + 5\bar t$, which is a color triplet and a fermion similar to
the quark of the fourth generation.

3. We have also considered ''b-replaced'' NBSs:
$T_S(n_bb-{\mbox{replaced}}) = n_b b + (6t + 6\bar t - n_b t)$ and
$T_f(n_bb-{\mbox{replaced}}) = n_b b + (6t + 5\bar t - n_b t),$
where $n_b$ is the integer number. The presence of b-quarks in the
NBS leads to the dominance of the isospin singlets: with the
inclusion of both b and t quarks we obtain a more weak isospin
invariant picture.

4. We have estimated the masses of the lightest ''b-replaced''
NBSs: $M_{T(b-{\mbox{replaced}})} \simeq (300 - 400)
\,\,{\mbox{GeV}},$ and predicted the existence of the more heavy
''b-replaced'' NBSs: $M_{T(n_bb-{\mbox{replaced}})} > 400
\,\,{\mbox{GeV}}$ with $n_b > 1$.

5. We have developed a theory of T-ball's condensate, and
predicted the possibility of the existence of three SM phases at
the EW-scale. Calculating the top-quark Yukawa coupling constant
at the border of two phases (with T-ball's condensate and without
it) we have obtained $g_t \approx 1$.

6. It was shown that CDF II Detector experiment searching for
heavy top-like quarks at the Tevatron (in $p\bar p$-collisions
with $\sqrt s \backsimeq 1.96$ GeV) can observe $T_f$-balls with
masses up to 400 GeV.

7. We have considered all processes with T-balls, which can be
observed at LHC, especially the decay $$H\to 2T_s$$ and the
production of $T_f\ov{T_f}$ as an alternative of the $t'\ov{t'}$
production (where $t'$ is the quark of the fourth generation with
t-quark quantum numbers).

8. We have constructed the possible form-factors of T-balls.

9. We have estimated the charge multiplicity (coming from the
T-ball's decays) at the energy W=14 TeV at LHC.

\section{Acknowledgments}

We deeply thank for the courtesy of CDF collaboration for the
presentation of figures from there.

H.B.N. is grateful to J.~Conway, R.~Erbacher, J.~Frost, H.~Jensen,
C.~Issever, E.~Lytken, K.~Loureiro and A.~Parker for advices and
fruitful discussions.

L.V.L. thanks A.B.~Kaidalov, O.V.~Pavlovsky and M.A.~Trusov for
useful discussions.

\end{document}